\documentclass[12pt]{article}

\usepackage{amsmath}

\newcommand{\eq}{\begin{equation}}
\newcommand{\eqx}{\end{equation}}
\newcommand{\eqs}{\begin{equation*}}
\newcommand{\eqsx}{\end{equation*}}
\newcommand{\eqn}{\begin{eqnarray}}
\newcommand{\eqnx}{\end{eqnarray}}
\newcommand{\alg}{\begin{align}}
\newcommand{\algx}{\end{align}}

\newcommand{\f}[2]{\frac{#1}{#2}}

\newcommand{\lm}{\lambda}

\renewcommand{\th}{\theta}

\newcommand{\slii}{{\tt sl(2)}\ }
\newcommand{\suii}{{\tt su(2)}\ }
\newcommand{\sutt}{{\tt su(2|2)}\ }

\DeclareMathOperator{\arcsinh}{arcsinh}

\newcommand{\tr}{\mbox{\rm tr}\,}
\newcommand{\nn}{{\cal N}}

\title{Four-loop perturbative Konishi from strings\\
and\\
finite size effects for multiparticle states
}

\author{Zoltan Bajnok$^{a}$\thanks{e-mail: {\tt bajnok@elte.hu}}\ \ and Romuald A. Janik$^{b}$\thanks{e-mail: {\tt ufrjanik@if.uj.edu.pl}}\\ \\
${}^a$ Theoretical Physics Research Group \\
       Hungarian Academy of Sciences \\
       1117 Budapest, P\'azm\'any s. 1/A \\
       Hungary\\
${}^b$ Institute of Physics\\
Jagellonian University,\\
ul. Reymonta 4, \\
30-059 Krak{\'o}w\\
Poland}

\begin{document}

\maketitle

\begin{abstract}
We derive the perturbative four loop anomalous dimension of the Konishi operator in $\nn=4$ SYM theory from the integrable string sigma model by evaluating the finite size effects using L{\"u}scher formulas adapted to multimagnon states at weak coupling. We obtain these multiparticle generalizations of L{\"u}scher formulas by studying certain exactly solvable relativistic integrable quantum field theories. The final result involves a summation of all bound states in the mirror theory and agrees with gauge theory perturbative computations.
\end{abstract}

\section{Introduction}

The AdS/CFT correspondence \cite{adscft} states the equivalence of $\nn=4$ Super-Yang-Mills gauge theory with superstrings on $AdS_5 \times S^5$. The correspondence is extremely interesting as it links the very difficult non-perturbative physics of gauge theory to (semi-)classical string/supergravity theory. As such it allows to gain new insight into various gauge theoretical phenomena but at the same time makes it very difficult to test and prove.

A real breakthrough in this respect is the discovery of integrability on both sides of the duality \cite{Minahan:2002ve,Beisert:2003tq,Beisert:2003yb,kor1,Bena:2003wd,Dolan:2003uh}. On the string theory side it means that the (light-cone quantized) worldsheet sigma model is an integrable quantum field theory, while on the gauge theory side it manifests itself in the appearance of spin chains. 

The question is to find the anomalous dimensions of all gauge theory operators as a function of the coupling $\lm=g^2_{YM}N_c$ or equivalently to find the quantized energy levels of a superstring in  $AdS_5 \times S^5$. Perturbative gauge theory physics corresponds to the deeply quantum regime of the worldsheet sigma-model.

Although worldsheet QFT and quantum spin chains seem to be completely different systems, their solution for large quantum numbers is encoded in the same mathematical structure -- the Bethe Ansatz \cite{BS0,KMMZ,K1,K2,BDS,S,BS}. In particular the S-matrix for the excitations is the same in both cases \cite{B} including the overall scalar factor (the so-called `dressing factor') which interpolates between strong and weak coupling, the absence of which was at the source of initial apparent disagreements between gauge and string theory. The specific form of the dressing factor is now believed to be known through a series of works \cite{AFS,T,CROSS,HL,BHL,BES} and the first nontrivial coefficient at weak coupling was checked by a direct perturbative evaluation \cite{BRM}.

In fact it is at the level of determining this scalar factor that the symmetry between the spin chain and worldsheet QFT picture is broken as the requirement of crossing symmetry \cite{CROSS} is definitely of a worldsheet QFT origin. A further difference from the spin-chain language arises since the Bethe ansatz quantization for a {\em quantum} field theory necessarily gets corrections due to virtual particles traveling around the cylinder \cite{Lusch1}. In \cite{AJK} it was suggested that thus necessarily the asymptotic Bethe Ansatz has to fail and that the natural magnitude of the corrections is consistent with the interpretation of these effects as wrapping interactions -- Feynman graphs which encompass the cylinder\footnote{The appearance of these graphs on the gauge theory side as a limitation to the results obtained from the asymptotic Bethe Ansatz was pointed out as early as \cite{BDS}.}. The main input of the 2D worldsheet QFT point of view was that in principle one expects these finite size effects to be {\em uniquely} determined from the infinite volume data. This is not so from the point of view of spin chains where it is far from obvious how to incorporate such effects, e.g. the Hubbard model was used \cite{Hubbard} but this could not be extended to the full theory with the dressing factor.

Deviations from the asymptotic Bethe ansatz have indeed been observed, mainly at strong coupling \cite{SS1,SS2,AFZ,Minahan} but also at weak coupling \cite{LSRV}. Some results at strong coupling were rederived \cite{JL,HS,SSV,HJL} using the formalism of L{\"u}scher corrections adapted to the nonrelativistic dispersion relation of the AdS magnons \cite{JL}. 

The calculation of finite size effects has another source of interest as it is sensitive to the details of the worldsheet theory as in particular {\em all} kinds of states are allowed to circulate in the loop. Thus questions about possible constituents of the fundamental magnons etc. could be tested using these methods. This is especially interesting at weak coupling as there one would be in a very `quantum' regime of the string sigma-model. Therefore it is extremely interesting to have perturbative results relevant to wrapping interactions.
Early work on the general structure of wrapping interactions in perturbative gauge theory was done in \cite{SIEG}.

The weak coupling result \cite{LSRV} is however a bit indirect and, for the moment, difficult to use, so various groups \cite{FSSZ,KM,Eden} set out to compute the anomalous dimension of the Konishi operator at four loop level -- the lowest order where these effects may appear. The details of the \cite{FSSZ} computation appeared in \cite{FSSZ2}.

The aim of this paper is to compute the same anomalous dimension of the Konishi operator at {\em gauge theory perturbative} four loop level starting from the worldsheet quantum field theory of the superstring in $AdS_5 \times S^5$ using its integrability properties. Thus the worldsheet QFT is defined in terms of the S-matrix which is a function of the 't Hooft coupling $\lm=g^2_{YM} N_c$. The leading three orders in $\lm$ of the anomalous dimensions follow from ordinary Bethe Ansatz quantization (which as described above mathematically coincides with the spin chain Bethe Ansatz). At order $\lm^4$, the effects due to virtual particles circulating around the cylinder start to play a role. Unfortunately, even in the relativistic case, direct analogs of L{\"u}scher formulas for multiparticle states (the Konishi operator corresponds to a two particle state in the spin-chain/QFT language) are not known. Thus we start from deriving such expressions in relativistic theories for which the finite size spectrum is known and use these results to conjecture their general form. Then we apply the resulting expressions to the concrete case of the two-particle state corresponding to the Konishi operator.

The plan of this paper is as follows. In section~2 we will briefly review the asymptotic Bethe ansatz description of the Konishi operator. In section~3 we will analyze relativistic integrable field theories and extract from them formulas for finite size corrections for multiparticle states. In section~4 we formulate our general conjecture for the appropriate formula which can be applied to the AdS case. Then in section~5, we describe all the ingredients needed for using the above formula for the case of the Konishi operator. In section~6 we derive the needed S-matrix elements between arbitrary bound states and the fundamental magnons. In section~7 we put together all the above ingredients and perform the computation for the Konishi operator. We close the paper with a discussion and several appendices.

\section{Konishi operator and the wrapping problem}

The Konishi operator is the simplest short non-protected operator in $\nn=4$ SYM which makes it a testing ground for the integrability approach to AdS/CFT. The operator is just
\eq
\tr \Phi_i^2
\eqx
where the $\Phi_i$'s are the six adjoint scalars. For the purposes of testing integrability it is more convenient to use two other operators which belong to the same super-multiplet and hence have the same anomalous dimension. One of these operators is the $L=4$ operator from the \suii sector $\tr X^2Z^2+\ldots$, and the other is an $L=2$ operator from the \slii sector $\tr D^2Z^2+\ldots$. The anomalous dimension can be evaluated using the asymptotic Bethe ansatz.
For definiteness let us consider the operator in the \suii sector. It corresponds to a $L=4$ spin chain with 2 excitations which we can take to have opposite momenta. The value of the momentum follows from the Bethe equation
\eq
e^{i4p}=\f{2u(p)+i}{2u(p)-i} e^{2i\th(p,-p)}
\eqx
where 
\eq
u(p)=\f{1}{2} \cot \f{p}{2} \sqrt{1+16g^2 \sin^2\f{p}{2}}
\eqx
and $e^{2i\th(p,-p)}$ is the dressing phase which, when evaluated to leading nontrivial order (for $p=2\pi/3$) is
\eq
e^{2i\th(p,-p)}=e^{-8\cdot 9\sqrt{3}\zeta(3)i g^6}
\eqx
The Bethe equation can be solved iteratively for the momentum
\eq
\label{e.p}
p=\f{2\pi}{3}-\sqrt{3}g^2+\f{9\sqrt{3}}{2} g^4 -\f{72\sqrt{3}+8\cdot 9\sqrt{3}\zeta(3)}{3} g^6+\ldots
\eqx
which when inserted into the magnon dispersion relation
\eq
E(p)=\sqrt{1+16g^2 \sin^2 \f{p}{2}}
\eqx
gives
\eq
\label{e.bethekon}
E_{Bethe}=4+12g^2-48g^4+336 g^6-(2820+288\zeta(3))g^8+\ldots
\eqx
This answer agrees with explicit perturbative computations up to three loops ($g^6$). At four loops qualitatively new types of perturbative contributions arise -- the so-called `wrapping interactions' which are not included in the asymptotic Bethe ansatz. 
The appearance of these new contributions make it impossible to proceed to strong coupling within the asymptotic Bethe ansatz for short operators.

We will isolate the wrapping interactions as deviations from the Bethe ansatz answer (\ref{e.bethekon}) thus we will split the total dimension of the Konishi operator as
\eq
E=E_{Bethe}+\Delta_{wrapping} E
\eqx
In the next two sections we derive the necessary formulas for computing leading finite size effects for multiparticle states and we will return to the Konishi operator in section~5.

\section{Finite size corrections for multiparticle states -- relativistic theories}

The leading finite size energy correction for excited states was first
analyzed by L\"uscher in \cite{Lusch1} using diagrammatic techniques. He concluded
that a standing one particle state acquires exponentially small corrections
in the volume. These corrections can be split into two parts: The
so called F-term includes the forward scattering amplitude: 
\eq
\Delta m_{F}(L)=-m\int_{-\infty}^{\infty}\frac{d\theta}{2\pi}\cosh\theta\,
\sum_{b}(S_{ab}^{ab}(\frac{i\pi}{2}+\theta)-1)e^{-mL\cosh\theta}
\eqx
while the $\mu$ term contains its residue: 
\eq
\Delta m_{\mu}(L)=- \sum_{b,c}\theta(m_{a}^{2}-\vert
m_{b}^{2}-m_{c}^{2}\vert)\mu_{ab}^c
(-i)\mbox{Res} S_{ab}^{ab}(\theta)\, e^{-\mu_{ab}^c L}
\eqx
where $\mu_{ab}^c$ is the altitude of the mass triangle with base $m_c $ 
if $c$ appears as a bound-state of $a$ and $b$,
otherwise zero. Contrary to this exponentially small corrections the
leading finite size effect of multiparticle states come from the
modification of the particles momenta $p_{a}=m_{a}\sinh\theta_{a}$.
It originates from consecutive scatterings on the other constituents:
\eq
e^{im_{a}\sinh\theta_{a}L}\prod_{b}S_{ab}^{ab}(\theta_{a}-\theta_{b})=
1\qquad;\qquad E(\{\theta_{a}\})=\sum m_{a}\cosh\theta_{a}
\eqx
These equations are called the Bethe-Yang (BY) equations and describe
the power-like corrections to the multiparticle energy levels exactly.
But of course, what we are really interested in are deviations from the result given by these Bethe-Yang equations.
Indeed there are exponentially small corrections as well and the
aim of this section to present a general form for their leading part.
A proper derivation would be to adopt L\"uscher's original diagrammatic
method, but since it is too involved and is beyond the scope of this
paper, we analyze certain exactly solvable relativistic quantum field 
theories for which the spectrum at finite size is known exactly.

We systematically
expand the excited states TBA equations of the sinh-Gordon model\footnote{This is technically easier than considering the SLYM model for which the excited state TBA was first derived in \cite{PatRob1} as there are no complications with bound states.} \cite{Jorg}
for large volumes from which we conjecture the general form of the
correction. This is then subject to various checks by comparing to
the excited states TBA of the Lee-Yang model on one side \cite{PatRob1} and to the
excited state NLIE of the sine-Gordon model on the other. Finally
we provide our conjecture for non-relativistically invariant theories
like AdS.

\subsection{Excited TBA equations, conjecture for the finite size correction }

In this part we recall how the finite volume energy levels of the
sinh-Gordon model can be described exactly following \cite{Jorg}.
The sinh-Gordon theory is one of the simplest integrable quantum field
theory. It contains one type of particle of mass $m$ whose factorized
scattering can be described in terms of the two particle scattering
matrix 
\eq
S(\theta)=\frac{\sinh\theta-i\sin B\pi}{\sinh\theta+i\sin B\pi}\qquad;\qquad B\geq0
\eqx
An $N$ particle state in volume $L$ is labeled by a set of integers
$\{n_{1},n_{2},\dots,n_{N}\}$ and its energy can be computed as follows:
First one has to solve the non-linear integral equation for the pseudo-energy
\eq
\epsilon(\theta)=mL\cosh\theta+\sum_{j=1}^{N}\log S(\theta-\theta_{j}-\frac{i\pi}{2})
-\int_{-\infty}^{\infty}\frac{d\theta^{'}}{2\pi}\phi(\theta-\theta^{'})
\log(1+e^{-\epsilon(\theta^{'})})   
\eqx
with
\eq
\phi(\theta)=\frac{d\log S(\theta)}{id\theta}
\eqx
where the particle rapidities $\theta_{j}$ are determined self-consistently
by the singularity $1+e^{-\epsilon(\theta_{j})}=0$ via the
prescribed integers $\{n_{j}\}$ as 
\eq
\epsilon(\theta_{j}+\frac{i\pi}{2})=i(2n_{j}+1)\pi\qquad;\qquad j=1,\dots,N
\eqx
Once the pseudo-energy and the rapidities are known the energy of
the multiparticle state is given by 
\eq
E_{\{n_{j}\}}(L)=m\sum_{j=1}^{N}\cosh\theta_{j}-m\int_{-\infty}^{\infty}
\frac{d\theta}{2\pi}\,\cosh\theta\,\log(1+e^{-\epsilon(\theta)})
\eqx
If we do not include any particles we recover the TBA equation of
the ground-state \cite{Alyosha}. We now turn to analyze the large
volume behaviour of the energy. For this we solve the TBA equation
by iteration and extract the leading finite size corrections.

\subsubsection*{Finite size correction for the ground-state}

Let us start with the ground-state. If $L$ is large we can approximate
the pseudo-energy by neglecting the exponentially small convolution
term as
\eq
\epsilon(\theta)=mL\cosh\theta+\dots
\eqx
By inserting this into the energy formula we obtain the leading finite size corrections
to the ground-state in the form 
\eq
E_{0}(L)=-m\int_{-\infty}^{\infty}\frac{d\theta}{2\pi}\,\cosh\theta\,
\log(1+e^{-mL\cosh\theta})\approx-m\int_{-\infty}^{\infty}\frac{d\theta}{2\pi}
\,\cosh\theta\, e^{-mL\cosh\theta}
\eqx

\subsubsection*{Finite size correction for a moving one-particle state}

A moving one-particle state is specified by its quantization number
$n$ from which its momentum can be calculated by solving the TBA
equation. In making the large volume expansion we have to be careful
and take into account one more iteration in order to
determine the pseudo-energy to the accuracy needed:
\eq
\epsilon(\theta)=mL\cosh\theta+\log S(\theta-\theta_{0}-\frac{i\pi}{2})
-\int_{-\infty}^{\infty}\frac{d\theta^{'}}{2\pi}\phi(\theta-\theta^{'})
S(\frac{i\pi}{2}+\theta_{0}-\theta^{'})e^{-mL\cosh\theta^{'}}
\eqx
The rapidity $\theta_{0}$ is fixed by $\epsilon(\theta_{0}+\frac{i\pi}{2})=i(2n+1)\pi$
which in this approximation reads
\eq
mL\sinh\theta_{0}+\delta\Phi=2i\pi n \ ;\quad\delta\Phi=
i\int_{-\infty}^{\infty}\frac{d\theta^{'}}{2\pi}\phi(\frac{i\pi}{2}
+{\theta_0}-\theta^{'})S(\frac{i\pi}{2}+{\theta_0}-\theta^{'})e^{-mL\cosh\theta^{'}}
\eqx
By looking for the solution in the form $\theta_{0}=\hat{\theta}_{n}+\delta\theta$
we can see that the free momentum quantization (the BY equation):
\eq
mL\sinh\hat{\theta}_{n}=2\pi n
\eqx
is also affected by the finiteness of the volume. The rapidity is
shifted by an exponentially small amount which reads 
\eq
\delta\theta=\frac{1}{\cosh\hat{\theta}_{n}}\int_{-\infty}^{\infty}
\frac{d\theta^{'}}{2\pi}\sinh\theta^{'}S(\frac{i\pi}{2}-\hat{\theta}_{n}
+\theta^{'})e^{-mL\cosh\theta^{'}}
\eqx
The energy correction comes from two places: from the modification
of the rapidity and also from the leading term in the pseudo-energy
\eq
E_{\{n\}}(L)=m\cosh\hat{\theta}_{n}+\delta\theta\, m\sinh\hat{\theta}_{n}
-m\int_{-\infty}^{\infty}\frac{d\theta}{2\pi}\,\cosh\theta\, S(\frac{i\pi}{2}
+\theta-\hat{\theta}_{n})e^{-mL\cosh\theta}
\eqx
Subtracting the contribution of the vacuum and using the explicit
form of $\delta\theta$ we can write 
\begin{eqnarray}
E_{\{n\}}(L)-E_{0}(L) & = & m\cosh\hat{\theta}_{n}\nonumber \\  &
&\hspace{-3cm}  -m\int_{-\infty}^{\infty} \frac{d\theta}{2\pi}\,
\f{\cosh(\theta-\hat{\theta}_{n})}{\cosh\hat{\theta}_{n}}\,
 \left(S(\frac{i\pi}{2}+\theta-\hat{\theta}_{n})-1\right)e^{-mL\cosh\theta}
\end{eqnarray}
Clearly for a standing particle $n=0$ there is no change in the momentum
$\theta_{0}=\hat{\theta}_{0}=0$ and we recover the F-term of the
L\"uscher correction. (Since in the sinh-Gordon theory there is no
bound-state the $\mu$ term is absent). Moreover, for generic states
we can confirm the result of \cite{KM} and \cite{JL}.

\subsubsection*{Finite size correction for multiparticle states -- diagonal scattering}

The pseudo-energy for a multiparticle state at the order needed can
be solved iteratively: 
\begin{eqnarray*}
\epsilon(\theta) & = & mL\cosh\theta+\sum_{j}\log
S(\theta-\theta_{j}-\frac{i\pi}{2})  \\
 &  & \int_{-\infty}^{\infty}\frac{d\theta^{'}}{2\pi}
\phi(\theta-\theta^{'})\prod_{j}S(\frac{i\pi}{2}+\theta_{j}-\theta^{'})
e^{-mL\cosh\theta^{'}}+\dots
\end{eqnarray*}
It gives rise to the quantization condition for each rapidity $\theta_{k}$
as 
\begin{align}
(2n_{k}+1)i\pi=&\epsilon(\theta_{k}+\frac{i\pi}{2})  =  
imL\sinh\theta_{k}+ i\pi+\sum_{j:j\neq k}  \log S(\theta_{k}-\theta_{j}) \nonumber\\
&-\int_{-\infty}^{\infty}\frac{d\theta^{'}}{2\pi}\phi(\frac{i\pi}{2}
+\theta_{k}- \theta^{'})\prod_{j}S(\frac{i\pi}{2}+\theta_{j}-\theta^{'})e^{-mL\cosh\theta^{'}}
\end{align}
The first part is the usual Bethe-Yang quantization condition 
\eq
2n_{k}\pi=BY_{k}(\theta_{1},\dots\theta_{n})=mL\sinh\theta_{k}-i\sum_{j:j\neq
  k} \log S(\theta_{k}-\theta_{j})
\label{eq:1pBY}
\eqx
while the second shows the effect of the non-trivial vacuum of the
finite volume and shifts the rapidities via modifying the BY equation:
\eqn
2n_{k}\pi&=&BY_{k}(\theta_{1},\dots\theta_{n})+\delta\Phi_{k} \nonumber\\
\delta\Phi_{k} &=& i\int_{-\infty}^{\infty}\frac{d\theta^{'}}{2\pi}
\phi(\frac{i\pi}{2} +\theta_{k}-\theta^{'})\prod_{j}S(\frac{i\pi}{2}
+\theta_{j}-\theta^{'})e^{-mL\cosh\theta^{'}}
\label{eq:1pBYmod}
\eqnx
Since $\delta\Phi_{k}$ is exponentially small we can parameterize
the rapidities as $\theta_{k}=\hat{\theta}_{k}+\delta\theta_{k}$
and express their change via 
\eq
\delta\theta_{k}=-\left(\frac{\delta BY_{k}}{\delta\theta_{j}}\right)^{-1}\delta\Phi_{j}
\eqx
Here $\left(\frac{\delta BY_{k}}{\delta\theta_{j}}\right)^{-1}$ is
the inverse of the matrix $\frac{\delta BY_{k}}{\delta\theta_{j}}$. 
The latter one is the Jacobi matrix of the change of variables between
the finite and infinite volume multiparticle basis. The full energy
correction comes from two places: from the shift of the rapidities
and also from the large volume asymptotic of the pseudo-energy:
\begin{align}
E(L)=&\sum_{k}m\cosh\hat{\theta}_{k}-\sum_{j,k}m\sinh\hat{\theta}_{k}
\left(\frac{\delta BY_{k}}{\delta\theta_{j}}\right)^{-1}\delta\Phi_{j}
+ \nonumber\\ &-m\int_{-\infty}^{\infty}\frac{d\theta}{2\pi}\,\cosh\theta\,
\prod_{k}S(\frac{i\pi}{2}+\theta-\hat{\theta}_{k})e^{-mL\cosh\theta}
\label{eq:1pEmod}
\end{align}

We conjecture the equations (\ref{eq:1pBYmod}-\ref{eq:1pEmod})
to be universal, valid in any two dimensional integrable field theory
with given scattering matrix. If the scattering matrix admits poles
in the physical strip corresponding to bound-states then we have to
sum the residues of the F-terms over the poles ($\theta^{*}$) of
the scattering matrices in the strip $0<\Im m(\theta^{*})<\frac{\pi}{2}$.
For instance in the Lee-Yang model, whose scattering matrix coincides
with the sinh-Gordon one for its non-physical value ($B=-\frac{2}{3}$),
we have such a pole. Then, additionally to the integral terms in 
eq.(\ref{eq:1pBYmod}) and eq.(\ref{eq:1pEmod}) we have extra terms
corresponding to the residues of the integrands. By expanding the
excited states TBA equations for the Lee-Yang model \cite{PatRob1}
in the same fashion as we did for the sinh-Gordon theory we were able
to confirm both eq.(\ref{eq:1pBYmod}) and eq.(\ref{eq:1pEmod})
together with the appropriate residue terms. 

The physical interpretation of the integral term in (\ref{eq:1pEmod})
is clear. A pair of particle and anti-particle appears from the vacuum,
they travel all over the world, scatter on the multiparticle state
and annihilate on the other side. This virtual process not only changes
the energy but also modifies the quantization condition, although not in 
an apparent way.  

It is very natural to extend our conjecture to theories with more species 
of particles and diagonal scattering. 
In such theories the Bethe quantization condition reads 
\begin{equation}
2n_{k}\pi=BY_{k}(\theta_{1},\dots\theta_{N})=m_{i_{k}}L\sinh\theta_{k}-i
\sum_{j:j\neq k}^{N}\log S_{i_{k}i_{j}}(\theta_{k}-\theta_{j})\label{eq:mpBYmod}
\end{equation}
in an obvious notation. Then following the philosophy of the sinh-Gordon model 
we expect these quantization conditions to be modified at leading order as
\begin{align}
2n_{k}\pi &=BY_{k}(\theta_{1},\dots\theta_{n})+\delta\Phi_{k} \nonumber\\ 
\delta\Phi_{k} &=\sum_{a=1}^{n}\int_{-\infty}^{\infty}\frac{d\theta^{'}}{2\pi}
\partial_{\theta_{k}}\log S_{ai_{k}}^{ai_{k}}(\frac{i\pi}{2}+\theta_{k}-\theta^{'})\prod_{j}
S_{ai_{j}}^{ai_{j}}(\frac{i\pi}{2}+\theta_{j}-\theta^{'})
e^{-mL\cosh\theta^{'}}
\label{eq:mpEmod}
\end{align}
where we summed up for all particles in the spectrum, since we expect
the appearance of any type of particle anti-particle pairs. We parameterized
the momenta as before $\theta_{k}=\hat{\theta}_{k}+\delta\theta_{k}$
and express $\delta\theta_{k}$ to leading order: $
\delta\theta_{k}=-\left(\frac{\delta
  BY_{k}}{\delta\theta_{j}}\right)^{-1}\delta\Phi_{j}
$.
 The full energy correction again comes from two places: from the shift
of the rapidities and also from the large volume asymptotic of the
pseudo-energy:
\begin{align}
E(L)&=\sum_{k}m_{i_{k}}\cosh\hat{\theta}_{k}-\sum_{j,k}m_{i_{k}}
\sinh\hat{\theta}_{k}\left(\frac{\delta BY_{k}}{\delta\theta_{j}}\right)^{-1}
\delta\Phi_{j}+ \nonumber\\ & -m_{a}\sum_{a}\int_{-\infty}^{\infty} 
\frac{d\theta}{2\pi}\,\cosh\theta\, \prod_{j}S_{ai_{j}}^{ai_{j}}
(\frac{i\pi}{2}+\theta-\hat{\theta}_{j})e^{-m_aL\cosh\theta}
\end{align}

This proposal can be subject to several checks in theories where the
exact energy levels are known. For instance we can analyze the generalizations
of the Lee-Yang model namely the $T_{n}$ systems which are perturbed
minimal models $\mathcal{M}_{2,2n+1}+\phi_{1,3}$ and for which excited
TBA equations are available \cite{PatRob2}. Indeed we checked by
expanding the excited state TBA equations of \cite{PatRob2} that
the F-terms which change the BY quantization and which changes the
energy are both correct. Moreover by shifting the contour $\theta\to\theta+\frac{i\pi}{2}$
we computed residues of the integrals. We observed that for a standing
particle the contributions corresponding to $\mu_{ab}^c $ and $\mu_{ac}^b$
cancel each other if $m_{a}^{2}<\vert m_{b}^{2}-m_{c}^{2}\vert$, and
adds up in the opposite case providing the correct $\mu$ term
in the L\"uscher formula. The physical picture one can associate
to this term, is that particle $a$ decays into particle $b$ and
$c$ which travel all over the world and then fuse back again to particle
$a$. Clearly in this picture both particles $b$ and $c$ have to
travel forward in time and this is exactly what is expressed by the
inequality $m_{a}^{2}>\vert m_{b}^{2}-m_{c}^{2}\vert$. 

There is a recent proposal for the $\mu$ -term (for symmetric fusion)
of the energy correction for multiparticle states \cite{Balazs}.
It consists of two terms, one describes the modification of the BY
quantization condition, while the other the correction to the energy.
By shifting the contour in our formulas we were able to recover the result
of \cite{Balazs} providing further evidence for both. 

Although in the sine-Gordon theory the scatterings are in general
non-diagonal we could find a subsector when our formulas can be tested.
In the repulsive regime, (when we have no bound-states), we can consider
an $N$ particle state built up purely from solitons. They scatter
diagonally on each other and are described by holes in the NLIE \cite{DdV,Gabor}.
By expanding the NLIE equation in a similar manner we did for the
sinh-Gordon TBA system we were able to confirm our formulas (\ref{eq:mpBYmod},\ref{eq:mpEmod})
for this special case. Moreover, by analyzing more complicated states
 we have gained some hints to the
generalization for non-diagonal scatterings.

\section{Finite size corrections for multiparticle states 
-- non-diagonal scattering and the AdS case
}
In this subsection we formulate our conjecture for non-diagonal scatterings.
First we suppose that the whole multiplet of particles has the same mass. 
The basic difference compared to the diagonal case is, that for an $N$
particle state labeled by the rapidities $\underline{
  \theta}=\{\theta_{1},\dots,\theta_{N}\}$ 
the consecutive
scatterings non only changes the momenta but also mixes the particle indexes 
$\underline{ a}=\{a_{1},\dots,a_{N}\}$.
Thus we have to diagonalize the multiparticle scattering matrices 
to obtain the quantization condition
\eq
e^{imL\sinh{\theta_k}}T(\theta_k\vert
\underline{ \theta})_{\underline{a}}^{\underline{ b}}
\Psi_{\underline{ b}} =e^{imL\sinh{\theta_k}}e^{i\delta (\theta_k
\vert{\underline{\theta}})}\Psi_{\underline{a}}=
\Psi_{\underline{ a}}\qquad \forall k
\eqx
Here we introduced a family of transfer matrices commuting
for different values of the auxiliary rapidity $\theta$:
\eq
T(\theta\vert {\underline{ \theta}})_{\underline{ a}}^{\underline{ b}}=(-1)^F
S_{c_1a_{1}}^{c_{2}b_{1}}(\theta,\theta_{1})S_{c_{2}a_{2}}^{c_{3}b_{2}}
(\theta,\theta_{2})\dots S_{c_{N}a_{N}}^{c_1b_{N}}(\theta,\theta_{N})
\eqx
which, thanks to unitarity $S_{ab}^{cd}(0)=(-1)^F\delta_a^c \delta_b^d$,
 reduces to the multiparticle scattering matrix for $\theta =\theta_k$. 
Here $F$ is the fermion-number operator. 
The matrices $T(\theta_{k}\vert {\underline{ \theta}})
_{\underline {a}}^{\underline{ b}}$ for different $\theta_{k}$ commute
with each other and 
each of their common eigenvalues corresponds to a multiparticle state. 
The rapidity of the particles
in the eigenstate $\Psi_{\underline{ a}}$ can be computed from
the equation 
\eq
2n_{k}\pi=mL\sinh\theta_{k}+\delta(\theta_{k}\vert{\underline{ \theta} })
=BY_{k}({\underline {\theta}})
\eqx
Now thanks to the experience we gained by analyzing the exact multiparticle
energy levels in the sine-Gordon theory together with the result of the
diagonal case we conjecture the modification of the BY equation to be
\eq
2n_{k}\pi=BY_{k}({\underline{ \theta}})+\delta\Phi_{k}\ ;\quad\delta\Phi_{k}
=\int_{-\infty}^{\infty}\frac{d\tilde{\theta }}{2\pi}\frac{d}{d\theta_{k}}
e^{i\delta(\tilde{\theta }\vert{\underline{ \theta} })}e^{-\tilde
  {\epsilon}(\tilde{\theta})L}
\eqx
Here we introduced the rapidity of the mirror theory $\tilde \theta=
\theta+\frac{i\pi}{2}$  and the mirror energy $\tilde \epsilon(\theta)=
-ip(\theta)=-im\sinh(\theta)$. See \cite{AJK} and \cite{AF1} for the details,
how the mirror theory appears from TBA type considerations. 
The exponentially small shifts of the rapidities turn out to be
$
\delta \theta_{k}=-\left(\frac{\delta BY_{k}}{\delta \theta_{j}}\right)^{-1}
\delta\Phi_{j}
$. Thus the full energy correction to the multiparticle energy 
$E=\sum_{k}\epsilon(\theta_{k})=\sum_km\cosh(\theta_k) $
can be written as:
\eq
E(L)=\sum_{k}\epsilon(\theta_{k})-\sum_{j,k}\frac{d\epsilon(\theta_k)}
{d\theta_{k}}
\left(\frac{\delta BY_{k}}{\delta
  \theta_{j}}\right)^{-1}\delta\Phi_{j}-\int_{-\infty}^{\infty}
\frac{d\tilde{\theta}}{2\pi}\frac{d\tilde p}{d\tilde \theta}
e^{i\delta(\tilde{\theta}\vert{\underline {\theta}})}
e^{-\tilde{\epsilon}(\tilde{\theta})L}
\eqx
where we used the mirror momentum $\tilde p=m\sinh \tilde{\theta}$. 

In generalizing to the AdS case we have to take into account its
non-relativistic behaviour and take care of the following 
properties: The scattering matrix depends separately on each of the momenta. 
The mirror theory whose excitations cause the correction both in BY and the 
energy are different from the theory itself, moreover we have an infinite
tower of bound-states of the fundamental particles.  
In view of these we formulate our 
conjecture for the simplest case when there is a type of particle
in the spectrum which  scatters on itself diagonally (which is the case for 
a magnon in one of the closed sectors say \suii or \slii). 
If we denote its label by $a$ then
the eigenvector of the transfer matrix is simply
$\Psi_{a\dots a}=1$ and all other elements are zero.
The eigenvalue of the transfer matrix is 
\eq
e^{i \delta(\tilde{p}\vert p_1,\dots ,p_N)} =(-1)^F\left[S_{a_{1}a}^{a_{2}a}(\tilde{p},p_{1})
S_{a_{2}a}^{a_{3}a}(\tilde{p},p_{2})\dots S_{a_{N}a}^{a_{1}a}(\tilde p,p_{N})\right]
\eqx 
As a consequence the BY condition reads as 
\eq
2n_{k}\pi=BY_{k}(p_{1},\dots p_{n})+\delta\Phi_{k}=p_{k}L-i\log
\left[\prod_{k\neq j}S_{aa}^{aa}(p_{k},p_{j})\right]+\delta\Phi_{k}
\eqx
We conjecture the correction to these equations to be 
\eq
\delta\Phi_{k}=-\int_{-\infty}^{\infty}\frac{d\tilde{p}}{2\pi}
(-1)^{F}
\left[S_{a_{1}a}^{a_{2}a}(\tilde{p},p_{1})\dots\frac{\partial
S_{a_{k}a}^{a_{k+1}a}(\tilde{p},p_{k})}{\partial\tilde{p}} 
\dots S_{a_{N}a}^{a_{1}a}(\tilde p,p_{N})\right]
e^{-\tilde{\epsilon_{a_1}}(\tilde{p})L}
\eqx
%The physical meaning of  $(-1)^F $ for fermion loops has the same origin, 
%why the ground-state
%energy corresponds to anti-periodic boundary condition for fermions in the
%time-like direction in the mirror theory. 
We also replaced
$\partial_{\theta_k}S(\tilde\theta -\theta_k)$ with
$-\partial_{\tilde\theta}S(\tilde\theta -\theta_k)$ before switching to $p$. With
this choice we could recover the one-particle result of \cite{JL}. 

The final correction then reads as 
\begin{align}
\label{e.fads}
E(L)&=\sum_{k}\epsilon(p_{k})-\sum_{j,k}\frac{d\epsilon(p_{k})}{dp_{k}} 
\left(\frac{\delta BY_{k}}{\delta p_{j}}\right)^{-1}\delta\Phi_{j} \nonumber\\ 
&-\int_{-\infty}^{\infty} \frac{d\tilde{p}}{2\pi} \sum_{a_1,\ldots,a_N}
(-1)^F  \left[ S_{a_{1}a}^{a_{2}a}(\tilde{p},p_{1})S_{a_{2}a}^{a_{3}a}
(\tilde{p},p_{2})\dots S_{a_{N}a}^{a_{1}a}(p,p_{N})\right]e^{-\tilde{\epsilon_{a_1}}(\tilde{p})L}
\end{align}
Here we explicitly wrote the summation to make clear, that what appears in 
the finite size correction is
in fact a supertrace and also that the indexes can change via the scatterings.
 
It seems plausible that by shifting the contour we can pick up the
$\mu$ terms of the correction. For related issues see also the derivation
of the $\mu$ term for a moving one particle state \cite{JL}. As
far as the physical interpretation is concerned we have to keep those
poles where the decayed particles travel forward in time, that is the 
real part of their energies are both positive. 

In \cite{PV} a proposal was presented for the multiparticle L\"uscher
correction based on a detailed analysis of the Hubbard model.
It is close in spirit to \cite{Balazs} since the authors analyzed the $\mu $
term coming from the composite structure of the particle. From this they conjectured the
finite size correction coming both from the modification of the Bethe Ansatz and from the
effect of the sea of particles and anti-particles. They formulated these corrections in
terms of the supertrace. The details in the general case are, however, somewhat different.
We conjecture our
formula to contain all the leading exponentially small corrections coming from single
particle contributions of the mirror theory. 

\section{Ingredients for the computation of the Konishi operator}

Let us now analyze the formula (\ref{e.fads}) from the point of view of its possible application to evaluate the leading wrapping corrections at weak coupling that appear at four-loop order. This might seem surprising at first glance as the Konishi is a very short operator of length $L=4$ (or even $L=2$ for the \slii representative). However as already advocated in \cite{AJK}, the specific dispersion relation of the magnons gives the estimate 
\eq
\label{e.naive}
e^{-2L \arcsinh \f{\sqrt{1+q^2}}{4g}} \rightarrow \f{4^L g^{2L}}{(1+q^2)^L}
\eqx
which for $L=4$ indeed gives the expected four-loop order $g^8$.
However those considerations did not control coupling constant dependence of the prefactor, as well as the precise definition of the length $L$ (equal to the circumference of the string worldsheet cylinder). In fact since the L{\"u}scher formula can be justified solely on the string theory side, because only then we are dealing with a 2D quantum field theory, the length $L$ that should be used is determined by the choice of light cone gauge for the string worldsheet. The simplest choice is to identify the length $L_{string}$ with $J$. Then since for the Konishi $J=2$ we have $L_{string}=2$ and from (\ref{e.naive}) we would have $g^4$. The remaining factors would then have to come from the $S$-matrix in the `string frame' \cite{ZF}. Thus even the {\em precise} leading order of the wrapping correction is not self-evident.

Another subtlety which one has to take into account is what states one should take to circulate in the loop. At strong coupling it is enough to consider just the fundamental magnons since the contribution of bound states is suppressed w.r.t the contribution of the fundamental magnons. This is not the case at weak coupling, were due to the dispersion relation
\eq
E_Q(p)=\sqrt{Q^2+16g^2 \sin^2 \f{p}{2}}
\eqx
the contribution analogous to (\ref{e.naive}) is
\eq
\f{4^J g^{2J}}{(Q^2+q^2)^J}
\eqx
So we have to sum over all $Q$. In order to perform the computation we have to know the full scattering matrix of all bound states with the fundamental magnon. This is known explicitly only in the case of the fundamental magnon ($Q=1$) and for the BPS bound states with $Q=2$ \cite{AF2}. In section 6 we will derive the general $S_{Q-1}$ matrix.

There is a final important problem which appears to be novel in the AdS case. It is well known that the physical bound states in the conventional theory are the ones appearing in the \suii sector \cite{Dorey1,Dorey2}. The potential bound states which would be associated to poles which appear in the \slii sector do not satisfy the physicality condition hence should not be considered as part of the spectrum. Yet when one considers the mirror theory (which exactly corresponds to the kinematic regime of the L{\"u}scher F-term) it turns out \cite{AF1} that it is the \slii poles which satisfy the physicality condition in the mirror theory. These states are quite different since they belong to different representations of \sutt -- the \suii bound states lie in the symmetric, while the \slii ones lie in the antisymmetric representation.
Hence we should decide which of the bound states should be taken into account when computing the L{\"u}scher F-term. Our conclusion is that one should take into account the \slii bound states. This is in fact quite natural from the TBA perspective, since in that approach one computes a partition function in the mirror theory\footnote{Although one has to keep in mind that such transparent physical picture exists only for ground state TBA.}.
In appendix D, for comparison we present the results of a corresponding computation using the \suii states. 

We will derive the S-matrix for scattering the antisymmetric states with the
fundamental one in the next section.

\section{S-matrix for symmetric and antisymmetric representations}

The S-matrix for the scattering of bound states with the fundamental magnons involves two basic ingredients: the matrix structure for all the polarization states of both particles and an overall scalar factor. The scalar factor can be fixed by specializing to the relevant closed subsector and fusing the bound state scattering from its elementary constituents (this has already been done for the \suii bound states \cite{DO}). Then the matrix structure can be found using the superfield formalism of \cite{AF2}. We will describe now both of these ingredients. 

We will use the notation $z^\pm$ for the bound state parameters satisfying
\eq
z^+ +\f{1}{z^+} -z^- -\f{1}{z^-}= \f{Qi}{g}
\eqx
while the parameters of the fundamental magnon will be denoted by $x^\pm$. They satisfy
\eq
x^+ +\f{1}{x^+} -x^- -\f{1}{x^-}= \f{i}{g}
\eqx

\subsection*{Scalar factors}

The overall scalar factor for bound state scattering in the \suii sector has
been found in \cite{DO,Roiban}. A convenient formula written directly in terms of the $z^\pm$ parameters of the bound state and $x^\pm$ parameters of the fundamental magnon was found\footnote{But note that the S-matrix there is the inverse of ours} in \cite{DO}:
\eq
\label{e.suiiscalar}
S_{Q-1}^{su(2)}(z^\pm,x^\pm)= \f{z^+-x^-}{z^--x^+} \f{1-\f{1}{z^+ x^-}}{1-\f{1}{z^- x^+}} \cdot \f{z^+-x^+}{z^--x^-} \f{1-\f{1}{z^+ x^+}}{1-\f{1}{z^- x^-}}
\eqx
This has to be supplemented by the dressing factor but it will not be important for our computations at weak coupling\footnote{Note however that due to the mirror kinematics of one of the two particles, the dressing factor scales as $\exp(\tilde{c} g^2)$ instead of $\exp(c g^6)$.}.

For the bound states of the mirror theory, the story is more complicated. It can be constructed completely algebraically by fusing the S-matrices of the scattering of the constituents of the \slii bound state $(z^-=z_1^-,z_1^+)$, $(z_2^-=z_1^+,z_2^+=z^+)$, etc. with the fundamental magnon $(x^-,x^+)$:
\eq
\label{e.sl2fuse}
S_{Q-1}^{sl(2)}(z^\pm,x^\pm)=\prod_{i=1}^Q S^{sl(2)}_{1-1}(z_i^\pm,x^\pm)
\eqx
with
\eq
S^{sl(2)}_{1-1}(z^\pm,x^\pm)=\f{z^--x^+}{z^+-x^-} \f{1-\f{1}{z^+ x^-}}{1-\f{1}{z^- x^+}}
\eqx
In contrast to the \suii case the result (\ref{e.sl2fuse}) depends on the
choice of the constituent magnons. Since the physical sheet is not known for 
the AdS S-matrix (see \cite{AF2} for related issues) we will adopt the choice
for which the constituents have the maximal number of parameters which scale
as $1/g$ and only the first $z_1^- \sim g$. This choice is consistent in the
sense that all but one of the constituents in the limit $g\to 0$ get close
to the real axis. So they are considered to be physical at the same time 
in the sense of \cite{Dorey2}. 
Depending on the choice of physical regions
\cite{AF1} identified  one, two or $2^{Q-1}$ boundstates in the spectrum. 
We used only the one mentioned above but it would be interesting to elaborate further the other choices, which might shed some light on the nature of the physical
domain. 
 
\subsection*{The matrix part}

Once we fix the scalar factor by the fusing relations for the closed sectors we have to normalize the matrix part of the S-matrix to be 1 for the `$11$' component in the case of the symmetric representation, and for the `$33$' component in the case of the antisymmetric one.

The aim of this part is to calculate the scattering matrix of the
fundamental magnon with the magnon of charge $Q$. In \cite{AF2}
this program was elaborated for the $Q=2$ case and symmetric representation corresponding to bound states in the \suii sector. Here we use their
conventions and extend their results first for arbitrary $Q$ and $1-Q$ scattering, and then we will obtain analogous results for bound states in the antisymmetric representation corresponding to the \slii sector\footnote{Recall that these are the {\em physical} bound states in the mirror theory.}. 

All of these particles belong to BPS representations of the supersymmetry
algebra $su(2\vert2)$. The algebra is defined by 
\begin{align*}
&[L_{a}^{b},J_{c}]=\delta_{c}^{b}J_{a}-\frac{1}{2}\delta_{a}^{b}J_{c}
  &[L_{a}^{b},J^{c}]=-\delta_{a}^{c}J^{b}+\frac{1}{2}\delta_{a}^{b}J^{c}\\{}
&[R_{\alpha}^{\beta},J_{\gamma}]=\delta_{\gamma}^{\beta}J_{\alpha}
-\frac{1}{2}\delta_{\alpha}^{\beta}J_{\gamma} &[R_{\alpha}^{\beta},J^{\gamma}]
=-\delta_{\alpha}^{\gamma}J^{\beta}+\frac{1}{2}\delta_{\alpha}^{\beta}J^{\gamma}\\
&\{Q_{\alpha}^{a},Q_{\beta}^{b}\}=\epsilon_{\alpha\beta}\epsilon^{ab}C\qquad 
&\{Q_{a}^{+\alpha},Q_{b}^{+\beta}\}=\epsilon_{ab}\epsilon^{\alpha\beta}C^{+}
\\ 
& & \{Q_{\alpha}^{a},Q_{b}^{+\beta}\}=\delta_{b}^{a}R_{\alpha}^{\beta}
+\delta_{\alpha}^{\beta}L_{b}^{a}+ 
\frac{1}{2}\delta_{b}^{a}\delta_{\alpha}^{\beta}H\hspace{2cm}
\end{align*}
where the central charges are expressed in terms of the momentum as
\eq
C=ig(e^{iP}-1)e^{2i\xi}\quad;\qquad C=-ig(e^{-iP}-1)e^{-2i\xi}\eqx
Two types of representations will be important: the atypical totally
symmetric and the totally anti-symmetric one. Both have dimension
$4Q$, and their specification for $Q=1$ give the fundamental representation. 

The totally symmetric representation can be realized by homogenous
symmetric polynomials of degree $Q$ of two bosonic ($w_{1},w_{2}$)
and two fermionic ($\theta_{3},\theta_{4}$) variables. The symmetry
operators on this space are represented by differential operators

\begin{eqnarray*}
L_{a}^{b}=w_{a}\frac{\partial}{\partial w_{b}}-\frac{1}{2}\delta_{a}^{b}w_{c}
\frac{\partial}{\partial w_{c}} & \qquad\qquad & R_{\alpha}^{\beta}
=\theta_{\alpha}\frac{\partial}{\partial\theta_{\beta}}-\frac{1}{2}
\delta_{\alpha}^{\beta}\theta_{\gamma}\frac{\partial}{\partial\theta_{\gamma}}\\
Q_{\alpha}^{a}=a\,\theta_{\alpha}\frac{\partial}{\partial w_{a}}
+b\epsilon^{ab}\epsilon_{\alpha\beta}w_{b}\frac{\partial}{\partial\theta_{\beta}}
 & \qquad\qquad & Q_{a}^{+\alpha}=d\,
w_{a}\frac{\partial}{\partial\theta_{\alpha}}
+c\epsilon^{\alpha\beta}\epsilon_{ab}\theta_{\beta}
\frac{\partial}{\partial w_{b}}
\end{eqnarray*}
 where 
\eq
a=\sqrt{\frac{g}{Q}}\eta\ ;\  b=\sqrt{\frac{g}{Q}}\frac{ie^{i2\xi}}
{\eta}(\frac{z^{+}}{z^{-}}-1)\ ;\  c=-\sqrt{\frac{g}{Q}}
\frac{\eta e^{-2i\xi}}{z^{+}}\ ;\  d=\sqrt{\frac{g}{Q}}
\frac{z^{+}}{i\eta}(1-\frac{z^{-}}{z^{+}})
\eqx
In accord with the previous notations in specializing the representation
for $Q=1$ we replace $z$ by $x$. 
There is a preferred choice for $\eta$ leading to unitary S-matrices given by \eq
\eta=e^{i\xi}e^{\frac{i}{4}p}\sqrt{iz^{-}-iz^{+}}
\label{eta}
\eqx
 The representations of the $su(2\vert2)$ algebra are characterized
by $\xi$ and $p$ and we denote the corresponding representation
by $\mathcal{V}^{Q}(p,e^{i2\xi})$. The scattering matrix acts as
\begin{equation}
S_{N-M}(p_{1},p_{2}):\quad\mathcal{V}^{N}(p_{1},e^{ip_{2}})
\otimes\mathcal{V}^{M}(p_{2},1)\longrightarrow\quad\mathcal{V}^{N}(p_{1},1)
\otimes\mathcal{V}^{M}(p_{2},e^{ip_{1}})\label{eq:SNM}
\end{equation}
 and commutes with the symmetry charges ($J$): 
\eq
S_{1-Q}(p_{1}p_{2})\left[J(p_{1},e^{ip_{2}})+J(p_{2},1)\right]
=\left[J(p_{1},1)+J(p_{2},e^{ip_{1}})\right]S_{1-Q}(p_{1},p_{2})
\eqx
Let us start from the $N=1$ and $M=Q$ case. Interestingly the
tensor product is irreducible 
\eq
\mathcal{V}^{1}\otimes\mathcal{V}^{Q}=\mathcal{W}^{Q+1}
\eqx
According to Schur's Lemma the invariance
fixes the scattering matrix up to a scalar factor. It is instructive
to use the $su(2)\otimes su(2)$ subalgebra generated by $L$ and
$R$ to parameterize the S-matrix. The representations in the tensor
product above decomposes as 
\begin{align*}
(V^{\frac{1}{2}}\otimes V^{0}&+V^{0}\otimes V^{\frac{1}{2}})
(V^{\frac{Q}{2}}\otimes V^{0}+V^{\frac{Q-1}{2}}\otimes V^{\frac{1}{2}}
+V^{\frac{Q-2}{2}}\otimes V^{0})  = \nonumber\\
&= V^{\frac{Q+1}{2}}\otimes V^{0}+3V^{\frac{Q-1}{2}}
\otimes V^{0}+2V^{\frac{Q}{2}}\otimes V^{\frac{1}{2}}+
  2V^{\frac{Q-2}{2}}\otimes V^{\frac{1}{2}} \nonumber\\
&\hspace{6cm}+V^{\frac{Q-1}{2}}\otimes V^{1}+V^{\frac{Q-3}{2}}\otimes V^{0}
\end{align*}
Since the S-matrix commutes with the $su(2)$ generators it has to
be in the form of $su(2)\otimes su(2)$ invariant differential operators
\eq
S_{1-Q}^{sym}=a_{1}^{1}\Lambda_{1}^{1}+\sum_{i,j=2}^{4}a_{i}^{j}\Lambda_{j}^{i}+
\sum_{i,j=5}^{6}a_{i}^{j}\Lambda_{j}^{i}+\sum_{i,j=7}^{8}a_{i}^{j}\Lambda_{j}^{i}+
a_{9}^{9}\Lambda_{9}^{9}+a_{10}^{10}\Lambda_{10}^{10}
\eqx
where the projectors are written in the form $\Lambda_{j}^{i}=v^{i}D_{j}$.
In Appendix~A we list the specific choice that we made.

As described in the beginning of this section our choice of 
overall scalar factor fixes the coefficient 
\eq
a_{1}^{1}=1
\eqx
The remaining coefficients are given in Appendix~B.

Since in the formula for the F-term we need the $S_{Q-1}$ and not the $S_{1-Q}$ S-matrix, we can use unitarity to relate the two matrices. Using eq. (3.34) in \cite{AF2} we find that this amounts to exchanging $x^+ \leftrightarrow x^-$ and $z^+ \leftrightarrow z^-$ in the formulas for the coefficients.

Now we turn to the analysis of the totally antisymmetric representations.
It can be realized by homogenous symmetric polynomials of degree $Q$
of two bosonic ($w_{3},w_{4}$) and two fermionic ($\theta_{1},\theta_{2}$)
variables. The symmetry charges act like the ones in the symmetric
representation except that we have to make the $w\leftrightarrow\theta$
replacement. The scattering matrix denoted as ${S}^{antisym}_{1-Q}({p}_{1},{p}_{2})$
can be parameterized as 
\eq
S^{antisym}_{1-Q}=\tilde{a}_{1}^{1}\tilde{\Lambda}_{1}^{1}+\sum_{i,j=2}^{4}
\tilde{a}_{i}^{j}\tilde{\Lambda}_{j}^{i}+ \sum_{i,j=5}^{6}\tilde{a}_{i}^{j}
\tilde{\Lambda}_{j}^{i}+
\sum_{i,j=7}^{8}\tilde{a}_{i}^{j}\tilde{\Lambda}_{j}^{i}
+\tilde{a}_{9}^{9} \tilde{\Lambda}_{9}^{9}+\tilde{a}_{10}^{10}\tilde{\Lambda}_{10}^{10}
\eqx
where $\tilde{\Lambda}$ can be obtained from $\Lambda$ by changing
the labels $1\leftrightarrow3$ and $2\leftrightarrow4$. 
This effectively interchanges the role of $Q$ and $Q^\dagger$ operators in the
previous computation so in the final formula one has essentially to
interchange $a$ with $d$ and $b$ with $c$. In order to see how this operation
acts on the coefficients expressed in terms of $x^\pm$ and $z^\pm$ one can use
the fact that $a=d^*$ and $b=c^*$, (with the choice made in (\ref{eta})), so
that this amounts to complex conjugation. Hence in the final formulas it will
be enough to interchange  $x^+ \leftrightarrow x^-$ and $z^+ \leftrightarrow
z^-$.

\section{The Konishi computation}

Let us now put together all the ingredients which enter the computation of
L{\"u}scher F-term adapted for the two particle state. The $\mu $ term is
absent in the weak coupling ($g\to 0$) limit, since the
fundamental particle with energy $\epsilon=1+\dots$ cannot decay into two
other particles with $\epsilon_{Q_1}=Q_1+\dots$ and
$\epsilon_{Q_2}=Q_2+\dots$ such that the charge is conserved $Q_1+Q_2=1$ and
the decayed particles propagate forward in time ($ \Re e (\epsilon_{Q_1})>0$
and $ \Re e (\epsilon_{Q_1})>0$). Thus we need only the F-term. 
 The terms responsible
for the modification of Bethe quantization conditions turn out to be of higher
order in $g$ therefore our formula takes the form 
\eq 
\Delta E_{wrapping}= \f{-1}{2\pi}\sum_{Q=1}^\infty \int_{-\infty}^\infty\! \! dq  \left(\f{z^-}{z^+}\right)^2 \sum_b (-1)^{F_b} \left[S_{Q-1}(z^\pm,x_i^\pm) S_{Q-1}(z^\pm,x_{ii}^\pm) \right]^{b(11)}_{b(11)}
\eqx
The summation over $Q$ is here over the fundamental magnons and all bound states of the mirror theory which are in the {\em antisymmetric} representation. The forward matrix element
\eq
\sum_b (-1)^{F_b} \left[S_{Q-1}(z^\pm,x_i^\pm) S_{Q-1}(z^\pm,x_{ii}^\pm) \right]^{b(11)}_{b(11)}
\eqx
is evaluated in terms of the coefficients of the S-matrices in Appendix~C.

The $x^\pm_i$ and $x^\pm_{ii}$ are the rapidities of the magnons constituting the Konishi operator. Their weak coupling expansion are obtained from
\eqn
x^\pm(p) &=& \f{1}{4g} \left(\cot \f{p}{2}\pm i \right) \left(1+ \sqrt{1+16g^2
  \sin^2 \f{p}{2}} \right) 
\eqnx 
so $x^\pm_i=x^\pm(p)$ and $x^\pm_{ii}=x^\pm(-p)$  with $p$ being the momentum (\ref{e.p}). To obtain the leading piece it is enough to use just the leading expression for the momentum $p=2\pi/3$.
The rapidity parameters of the mirror bound states follow from expanding
\eq
z^\pm = \f{Q}{4g} \left( -\sqrt{1+\f{16g^2}{Q^2+q^2}}\mp 1\right) \left(-\f{q}{Q}-I\right)
\eqx
Using these expressions we find immediately the exponential piece:
\eq
\left(\f{z^-}{z^+}\right)^2= \f{16 g^4}{(Q^2+q^2)^2}+\ldots
\eqx
The forward matrix element factorizes into the scalar part and a matrix part.

The scalar part is obtained by fusion from (\ref{e.sl2fuse}) using for $z_1^-$ the value $z^-$ above, taking $z_Q^+$ to be equal to $z^+$, and calculating $z_k^+$ from
\eq
z_k^+=\f{1}{2} \left( z_k^-+\f{1}{z_k^-} +\f{i}{g}+ \sqrt{\left( z_k^-+\f{1}{z_k^-} +\f{i}{g} \right)^2-4 } \right)
\eqx
The intermediate $z_k^-$ are determined from the pole condition $z_{k}^-=z_{k-1}^+$. The final result for the scalar part is
\begin{align}
S^{scalar,sl(2)}_{Q-1}=&
\frac{3 q^2-6 i Q q+6 i q-3 Q^2+6 Q-4}{3 q^2+6 i Q q-6 i
   q-3 Q^2+6 Q-4} \cdot \nonumber\\
& \frac{16}{9 q^4+6 (3 Q (Q+2)+2) q^2+(3 Q (Q+2)+4)^2}
\end{align}
The matrix part can be evaluated using the formulas (\ref{FSA}-\ref{F2}). The result is
\eq  
S^{matrix,sl(2)}_{Q-1}= \f{5184 Q^2 (3q^2+3Q^2-4)^2 g^4}{(q^2+Q^2)^2 ((3q-3iQ+3i)^2-3)^2}
\eqx
Let us note some features of this result. Firstly, it is of order $g^4$ so effectively the length is increased from $J=2$ to $L=4$. Secondly, although we are evaluating the leading behaviour at weak coupling we cannot use the 1-loop S-matrix as it stands since the mirror particle has $z^- \sim g$ and $z^+ \sim 1/g$ which modifies some pieces of the S-matrix. Thirdly, although the S-matrix has to be derived separately for $Q=1$, $Q=2$ and $Q\geq 3$, the forward S-matrix element has a uniform expression as a rational function of $Q$.

Putting the above expressions together we obtain
\begin{align}
\f{-1}{2\pi} \cdot g^8 \cdot & 
\frac{147456  Q^2 \left(3 q^2+3
   Q^2-4\right)^2}{\left(q^2+Q^2\right)^4 \left(9 q^4+6
   (3 (Q-2) Q+2) q^2+(3 (Q-2) Q+4)^2\right)} \cdot \nonumber\\
& \frac{1}{9 q^4+6 (3 Q (Q+2)+2) q^2+(3 Q (Q+2)+4)^2}
\end{align}
which has to be integrated over the real line of $q$ and summed over $Q$. One can easily evaluate the integral by residues. It is convenient to present the result using a partial fraction expansion for a part of the answer:
\eq
\label{e.sumQ}
-\frac{num(Q)}{\left(9 Q^4-3
   Q^2+1\right)^4 \left(27 Q^6-27 Q^4+36 Q^2+16\right)}
+\frac{864}{Q^3}-\frac{1440}{Q^5} 
\eqx
where the numerator $num(Q)$ is given by
\begin{align}
num(Q)=& 7776 Q (19683 Q^{18}-78732 Q^{16}+150903
   Q^{14}-134865 Q^{12}+ \nonumber\\
&+1458 Q^{10}+48357 Q^8-13311
   Q^6-1053 Q^4+369 Q^2-10)
\end{align}
We now have to sum eq.(\ref{e.sumQ}) over $Q$ running from 1 to $\infty$. The two last terms directly give $864\zeta(3)-1440 \zeta(5)$, while the remaining complicated rational function sums up to an integer. So the final answer for the wrapping correction at 4-loop coming from the finite size effects in the worldsheet string QFT is
\eq
\Delta_{wrapping}E=(324+864\zeta(3)-1440 \zeta(5))g^8
\eqx
Let us emphasize that the fact that such a simple transcendentality structure appeared is far from obvious. The key point is that the very complicated rational function part of (\ref{e.sumQ}) sums up exactly to an {\em integer}. In appendix D we describe a similar computation using the conventional `\suii' bound states in the symmetric representation and there an analogous rational function of $Q$ generates a lot of $\psi$ functions, $\pi^2$ which do not seem to simplify. 

Let us compare our result with the result of ref. \cite{FSSZ2}. Their result expressed as a correction to the asymptotic Bethe result is\footnote{After this paper appeared, the integer part of the result of \cite{FSSZ2} was corrected giving now exact agreement.}
\eq
\label{e.zanon}
\Delta_{wrapping}^{\mbox{\small ref.\cite{FSSZ2}}}E=(324+864\zeta(3)-1440 \zeta(5))g^8
\eqx
thus giving an exact agreement between the gauge theory perturbative computation of \cite{FSSZ2} and the worldsheet string sigma model computation of the present paper.
We will comment on its possible implications in the Discussion.

\section{Discussion}

In this paper we have derived generalizations of L{\"u}scher formulas for multiparticle states by examining the structures appearing in integrable relativistic theories solvable in finite volume using the techniques of the Thermodynamic Bethe Ansatz (TBA). The novel feature which appears is the modification of Bethe quantization conditions which effectively changes in a slightly nonintuitive way the formula for the `F-term'. The counterpart of the `$\mu$-term' follows by taking residues.

Then based on the intuition gained from the relativistic formulas, as well as comparison with generalized L{\"u}scher formulas for single particle states derived in \cite{JL}, we proposed a generalization of these multiparticle L{\"u}scher formulas to the AdS case.

Using the proposed formulas we considered the Konishi operator and derived the four-loop wrapping contribution to the anomalous dimension. In doing so we had to sum over an infinite set of bound states {\em of the mirror theory} which lie in the antisymmetric representation and correspond to the \slii sector. This is rather surprising since these states are not physical for the ordinary theory where the symmetric bound states of the \suii are the physical ones. They are however physical states of the mirror theory \cite{AF1}. In order to carry out the computation we have derived the full S-matrix between the $Q$-bound state and a fundamental magnon, since in computing the F-term correction we have to sum over all `polarization' states of the bound states.

The resulting expression for the F-term contribution is
\eq 
\Delta_{wrapping}E=(324+864\zeta(3)-1440 \zeta(5))g^8
\eqx
leading to the total anomalous dimension of the Konishi operator
\eq
E=4+12g^2-48g^4+336 g^6+(-2496+576\zeta(3)-1440\zeta(5))g^8+\ldots
\eqx
This result has the transcendentality structure expected from perturbative gauge theory point of view. Let us note that this is a very stringent test on the string theory result as generically e.g. when looking at poles or using the \suii family of bound states (see appendix D) one obtains much more complicated expressions which are not expected from the point of view of a gauge theory perturbative computation.

Our result coincides with the direct four-loop gauge theoretical computation of \cite{FSSZ2} quoted in (\ref{e.zanon}). Since the finite size effects are sensitive to all possible particles of the theory which circulate in the virtual loop, and moreover are very subtle at weak coupling where all bound states contribute equally, they are a very sensitive test of our knowledge of the details of the worldsheet string theory.
The exact agreement is a very strong indication that the description of the light cone quantized superstring in $AdS_5 \times S^5$ in terms of magnons and bound states is complete.

Moreover let us emphasize that the computation of the finite size corrections used in an essential way the language of two dimensional {\em quantum field theory} in order to describe weak coupling perturbative effects. The complete agreement with gauge theory is a very strong indication of the validity of the continuum worldsheet string theory description of these phenomena.
%Thus it would be very interesting to reconfirm the gauge theory perturbative computation.

\bigskip

\noindent{}{\bf Acknowledgments.} We thank Sergei Frolov and Radu Roiban for
valuable insight on bound states and perturbative expectations and \'Arp\'ad
Heged{\H u}s,  L\'aszl\'o
Palla, J\"org Teschner and Changrim Ahn for useful discussions. 
 RJ thanks the Galileo Galilei Institute for
Theoretical Physics for the hospitality and the INFN for partial support
during the completion of this work.
This work has been supported in part by Polish Ministry of Science and
Information Technologies grant 1P03B04029 (2005-2008), RTN network
ENRAGE MRTN-CT-2004-005616, and the Marie Curie ToK KraGeoMP (SPB
189/6.PRUE/2007/7). ZB was supported by a Bolyai Scolarship, OTKA 60040 and
the EC network ``Superstring''. 

\appendix

\section{Basis of projectors for the S-matrix computation}
The $su(2)\otimes su(2)$ invariant projectors $\Lambda_i^j $ can be written in
the form $\Lambda_i^j =v_iD^j $  where summation over the repeated indexes is
understood.  We made the following choices in parameterizing the scattering
matrix: 
\eqs
v_{1}= \frac{1}{Q+1}(w_{a}^{1}w_{a_{1}}^{2}\dots w_{a_{Q}}^{2}+w_{a}^{2}w_{a_{1}}^{1}
\dots w_{a_{Q}}^{2}+\dots+w_{a}^{2}w_{a_{1}}^{2}\dots w_{a_{Q}}^{1})
\eqsx
\eqs
D^{1}= \frac{1}{Q!}\frac{\partial^{Q+1}}{\partial w_{a}^{1}\partial
  w_{a_{1}}^{2} \dots\partial w_{a_{Q}}^{2}}
\eqsx
\eqs
v_{2}=  \frac{1}{2}(w_{1}^{1}w_{2}^{2}-w_{2}^{1}w_{1}^{2})w_{a_{1}}^{2}\dots
w_{a_{Q-1}}^{2}
\eqsx
\eqs
D^{2}= \left[\frac{\partial^{2}}{\partial w_{1}^{1}\partial w_{2}^{2}}
-\frac{\partial^{2}}{\partial w_{2}^{1}\partial w_{1}^{2}}\right]
\frac{1}{(Q-1)!}\frac{\partial^{Q-1}}{\partial w_{a_{1}}^{2}\dots
\partial w_{a_{Q-1}}^{2}}
\eqsx
\eqs
v_{3}= \frac{1}{2}(\theta_{3}^{1}\theta_{4}^{2}-\theta_{4}^{1}\theta_{3}^{2})
w_{a_{1}}^{2}\dots w_{a_{Q-1}}^{2}
\eqsx
\eqs
D^{3}=\left[\frac{\partial^{2}}{\partial\theta_{4}^{2}\partial\theta_{3}^{1}}
-\frac{\partial^{2}}{\partial\theta_{3}^{2}\partial\theta_{4}^{1}}\right]
\frac{1}{(Q-1)!}\frac{\partial^{Q-1}}{\partial w_{a_{1}}^{2}\dots
\partial w_{a_{Q-1}}^{2}}
\eqsx
\eqs
v_{4}= \frac{1}{Q-1}(w_{a_1}^{1}w_{a_{2}}^{2}\dots w_{a_{Q-1}}^{2}+\dots+
w_{a_1}^{2}w_{a_{2}}^{2}\dots w_{a_{Q-1}}^{1})\theta_{3}^{2}\theta_{4}^{2}
\eqsx
\eqs
D^{4}= \frac{2}{(Q-1)!}\frac{\partial^{Q-1}}{\partial w_{a_1}^{1}\partial 
w_{a_{2}}^{2}\dots\partial w_{a_{Q-1}}^{2}}\frac{\partial^{2}}
{\partial\theta_{4}^{2}\partial\theta_{3}^{2}}
\eqsx
\eqs
v_{5}=  \theta_{\alpha}^{1}w_{a_{1}}^{2}\dots w_{a_{Q}}^{2}
\eqsx
\eqs
D^{5}= \frac{\partial}{\partial\theta_{\alpha}^{1}}\,\frac{1}{Q!}
\frac{\partial^{Q}}{\partial w_{a_{1}}^{2}\dots\partial w_{a_{Q}}^{2}}
\eqsx
\eqs
v_{6}=\frac{1}{Q}(w_{a_1}^{1}w_{a_{2}}^{2}\dots w_{a_{Q}}^{2}+\dots+w_{a_1}^{2}
w_{a_{2}}^{2}\dots w_{a_{Q}}^{1})\theta_{\alpha}^{2}
\eqsx
\eqs
D^{6}= \frac{1}{Q!}\frac{\partial^{Q}}{\partial w_{a_1}^{1}\partial
  w_{a_{2}}^{2} \dots\partial w_{a_{Q}}^{2}}\,\frac{\partial}
{\partial\theta_{\alpha}^{2}}
\eqsx
\eqs
v_{7}= \frac{1}{2}(w_{1}^{1}w_{2}^{2}-w_{2}^{1}w_{1}^{2})w_{a_{1}}^{2}
\dots w_{a_{Q-2}}^{2}\theta_{\alpha}^{2}
\eqsx
\eqs
D^{7}= \left[\frac{\partial^{2}}{\partial w_{1}^{1}\partial w_{2}^{2}}
-\frac{\partial^{2}}{\partial w_{2}^{1}\partial w_{1}^{2}}\right]
\frac{1}{(Q-2)!}\frac{\partial^{Q-2}}{\partial w_{a_{1}}^{2}\dots
\partial w_{a_{Q-2}}^{2}}\frac{\partial}{\partial\theta_{\alpha}^{2}}
\eqsx
\eqs
v_{8}= \theta_{\alpha}^{1}w_{a_{1}}^{2}\dots
w_{a_{Q-2}}^{2}\theta_{3}^{2}\theta_{4}^{2}
\eqsx
\eqs
D^{8}= \frac{\partial}{\partial\theta_{\alpha}^{1}}\frac{2}{(Q-2)!}
\frac{\partial^{Q-2}}{\partial w_{a_{1}}^{2}\dots\partial w_{a_{Q-2}}^{2}}
\frac{\partial^{2}}{\partial\theta_{4}^{2}\partial\theta_{3}^{2}}
\eqsx
\eqs
v_{9}= \frac{1}{2}(\theta_{\alpha}^{1}\theta_{\alpha_{1}}^{2}
+\theta_{\alpha_{1}}^{1}\theta_{\alpha}^{2})w_{a_{1}}^{2}\dots w_{a_{Q-1}}^{2}
\eqsx
\eqs
D^{9}= \frac{\partial^{2}}{\partial\theta_{\alpha_{1}}^{2}\partial
\theta_{\alpha}^{1}}\frac{1}{(Q-1)!}\frac{\partial^{Q-1}}
{\partial w_{a_{1}}^{2}\dots\partial w_{a_{Q-1}}^{2}}
\eqsx
\eqs
v_{10}= \frac{1}{2}(w_{1}^{1}w_{2}^{2}-w_{2}^{1}w_{1}^{2})w_{a_{1}}^{2}
\dots w_{a_{Q-3}}^{2}\theta_{3}^{2}\theta_{4}^{2}
\eqsx
\eqs
D^{10}= \left[\frac{\partial^{2}}{\partial w_{1}^{1}\partial w_{2}^{2}}
-\frac{\partial^{2}}{\partial w_{2}^{1}\partial w_{1}^{2}}\right]
\frac{1}{(Q-3)!}\frac{\partial^{Q-3}}{\partial w_{a_{1}}^{2}\dots
\partial w_{a_{Q-3}}^{2}}\frac{\partial^{2}}{\partial\theta_{4}^{2}
\partial\theta_{3}^{2}}
\eqsx
These projectors are symmetric $(\Lambda_i^j)^\dagger =\Lambda_j^i$ for the 
$(w_a^i)^\dagger=\frac{\partial}{\partial w_a^i}$ and  $(\theta_{\alpha}^i)^\dagger=
\frac{\partial}{\partial \theta_{\alpha}^i}$ transformations. As a consequence
  the scattering matrix is also symmetric in terms of the coefficients $a$. 

\section{Coefficients of the $S_{1-Q}$ S-matrix for the symmetric representation}

We calculated the elements of the $S_{1-Q}$  scattering matrix by demanding
its commutation with the conserved charges and obtained the following result:
\eqs
a_{5}^{5}=\frac{x^{+}-z^{+}}{x^{+}-z^{-}}\frac{\tilde{\eta}_{1}}
{\eta_{1}}
\eqsx
\eqs
a_{5}^{6}=\sqrt{Q}\frac{(x^{+}-x^{-})}{(x^{+}-z^{-})}
\frac{\tilde{\eta}_{2}}{\eta_{1}}\quad;\qquad a_{6}^{5}=\sqrt{Q}
\frac{(z^{+}-z^{-})}{(x^{+}-z^{-})}\frac{\tilde{\eta}_{1}}{\eta_{2}}
\eqsx
\eqs
a_{6}^{6}=Q\frac{x^{-}-z^{-}}{x^{+}-z^{-}}
\frac{\tilde{\eta}_{2}}{\eta_{2}}
\eqsx
\eqs
a_{9}^{9}=\frac{x^{-}-z^{+}}{x^{+}-z^{-}}
\frac{\tilde{\eta}_{1}}{\eta_{1}}\frac{\tilde{\eta}_{2}}{\eta_{2}}\eqsx
\eqs
a_{7}^{7}=\frac{2}{Q}\frac{z^{-}(x^{-}-z^{+})(1-x^{-}z^{+})}
{z^{+}(x^{+}-z^{-})(1-x^{-}z^{-})}\frac{\tilde{\eta}_{2}}{\eta_{2}}
\eqsx
\eqs
a_{8}^{8}=\frac{x^{-}(x^{-}-z^{+})(1-x^{+}z^{-})}{2x^{+}(x^{+}-z^{-})
(1-x^{-}z^{-})}\frac{\tilde{\eta}_{1}}{\eta_{1}}\frac{\tilde{\eta}_{2}^{2}}
{\eta_{2}^{2}}
\eqsx
\eqs
a_{10}^{10}=\frac{2}{Q-1}\frac{x^{-}z^{-}(x^{-}-z^{+})(1-x^{+}z^{+})}
{x^{+}z^{+}(x^{+}-z^{-})(1-x^{-}z^{-})}\frac{\tilde{\eta}_{2}^{2}}{\eta_{2}^{2}}
\eqsx
\eqs
a_{7}^{8}=-\frac{i}{\sqrt{Q}}\frac{x^{-}z^{-}(x^{-}-z^{+})}{x^{+}z^{+}
(x^{+}-z^{-})(1-x^{-}z^{-})}\frac{\tilde{\eta}_{1}\tilde{\eta}_{2}^{2}}{\eta_{2}}
\eqsx
\eqs a_{8}^{7}=\frac{i}{\sqrt{Q}}\frac{(x^{-}-x^{+})(x^{-}-z^{+})
(z^{-}-z^{+})}{(x^{+}-z^{-})(1-x^{-}z^{-})}\frac{\tilde{\eta}_{2}}
{\tilde{\eta}_{1}\eta_{2}^{2}}\eqsx
\eqs
a_{2}^{3}=\frac{2i}{\sqrt{Q}}\frac{x^{-}z^{-}(x^{+}-z^{+})\tilde{\eta}_{1}
\tilde{\eta}_{2}}{x^{+}z^{+}(x^{+}-z^{-})(1-x^{-}z^{-})}
\eqsx
\eqs
a_{3}^{2}=-\frac{2i}{\sqrt{Q}}\frac{(x^{-}-x^{+})(z^{-}-z^{+})(x^{+}-z^{+})}
{x^{+}z^{+}(x^{+}-z^{-})(1-x^{-}z^{-})\eta_{1}\eta_{2}}
\eqsx
\eqs
a_{2}^{4}=-i\frac{Q-1}{Q}\frac{x^{-}z^{-}(x^{-}-x^{+})
\tilde{\eta}_{2}^{2}}{x^{+}z^{+}(x^{+}-z^{-})(1-x^{-}z^{-})}
\eqsx
\eqs
 a_{4}^{2}=i\frac{Q-1}{Q}\frac{(z^{-}-z^{+})^{2}(x^{-}-x^{+})}
{(x^{+}-z^{-})(1-x^{-}z^{-})\eta_{2}^{2}}
\eqsx
\eqs
a_{3}^{4}=\frac{Q-1}{\sqrt{Q}}\frac{x^{-}(x^{+}-x^{-})(1-x^{+}z^{-})}
{x^{+}(x^{+}-z^{-})(1-x^{-}z^{-})}\frac{\tilde{\eta}_{2}^{2}}{\eta_{1}\eta_{2}}\eqsx
\eqs a_{4}^{3}=\frac{Q-1}{\sqrt{Q}}\frac{x^{-}(z^{+}-z^{-})(1-x^{+}z^{-})}
{x^{+}(x^{+}-z^{-})(1-x^{-}z^{-})}\frac{\tilde{\eta}_{1}\tilde{\eta}_{2}}{\eta_{2}^{2}}
\eqsx
\begin{eqnarray*}
a_{2}^{2}=-\frac{1}{Q(1+Q)}\frac{1}{z^{+}(x^{+}-z^{-})(1-x^{-}z^{-})}\left [
  2z^{-}z^{+}(Q+x^{-}z^{-}-(1+Q)x^{-}z^{+}) \right .\nonumber \\
\left . +2x^{+}(z^{+}+z^{-}(-1+Q(-1+x^{-}z^{+}))) \right ]
\end{eqnarray*}
\eqs
a_{3}^{3}=\frac{(-x^{-}x^{+}(1+x^{-}z^{-}-2x^{+}z^{-})-(x^{+}+x^{-}(-2+x^{+}z^{-}))
z^{+})}{x^{+}(x^{+}-z^{-})(1-x^{-}z^{-})}\frac{\tilde{\eta}_{1}}{\eta_{1}}
\frac{\tilde{\eta}_{2}}{\eta_{2}}
\eqsx
\begin{eqnarray*}
a_{4}^{4}=-\frac{(Q-1)}{2Q(x^{+})^{2}
(x^{+}-z^{-})(1-x^{-}z^{-})}\left [
  x^{-}(Q((x^{-})^{2}x^{+}z^{-}-x^{-}(x^{+}+z^{-})
\right . \nonumber \\
\left . +x^{+}z^{-}(2-x^{+}z^{-}))-(x^{-}-x^{+})x^{+}z^{-}(z^{-}-z^{+}))
\right ] \frac{\tilde{\eta}_{2}^{2}}{\eta_{2}^{2}}
\end{eqnarray*}

Here the parameters $\eta$ and $\tilde \eta$ are representing a freedom in
choosing the basis in the fermionic generators. With the choice   
\eq
\eta_{1}=e^{ip_{2}/2}\eta(x)\:,\quad\eta_{2}=\eta(z)\:,\quad\tilde{\eta}_{1}
=\eta(x)\:,\quad\tilde{\eta}_{2}=e^{ip_{1}/2}\eta(z)
\eqx
the scattering matrix is symmetric. 

\section{Forward scattering element for general $Q$}

In this Appendix we express the forward scattering elements in terms of the
scattering matrix we determined in the previous appendix. We focus on the
symmetric representation and the $1-Q$ scattering of the $su(2)$ sector first.
 Recall that the forward scattering element can be written
\eq
FSA_{1-Q}^{sym}(x_i^\pm,x_{ii}^\pm,z^\pm)=\sum_b (-1)^{F_b} \left[S_{1-Q}(x_i^\pm,z^\pm) S_{1-Q}(x_{ii}^\pm,z^\pm) 
\right]^{(11)b}_{(11)b}
\eqx
Each individual scattering matrix has the form 
\eq
S_{1-Q}(x^\pm,z^\pm)_{(11)(a\dot a)}^{(11)(b\dot b)}=S^{su(2)}_{1-Q}(x^\pm,z^\pm)
S^{sym}_{1-Q}(x^\pm,z^\pm)_{1a}^{1b}S^{sym}_{1-Q}(x^\pm,z^\pm
)_{1\dot a}^{1\dot b}
\eqx
where we used the $su(2\vert 2)\otimes su(2\vert 2)$
factorization of the problem.  Due to this factorized form the forward
scattering can be written as 
\begin{eqnarray}
FSA_{1-Q}^{sym}(x_i^\pm,x_{ii}^\pm,z^\pm)
=S^{scalar,su(2)}_{1-Q}S^{matrix,su(2)}_{1-Q}=\hspace{4cm} \\ 
S^{su(2)}_{1-Q}(x_i^\pm,z^\pm)S^{su(2)}_{1-Q}(x_{ii}^\pm,z^\pm)
\left ( \sum_{a,b=1}^{4Q} (-1)^F S^{sym}_{1-Q}(x_i^\pm,z^\pm)_{1a}^{1b}
S^{sym}_{1-Q}(x_{ii}^\pm,z^\pm)_{1b}^{1a}\right )^2 \nonumber
\label{FSA}
\end{eqnarray}
We choose the basis in the $4Q$ dimensional bound-state representation as
follows: For the range of the index  $a=1,\dots ,Q+1$ we associate a state 
 $w_1^1w_1^2 \dots w_1^2w_2^2 \dots w_2^2 $ with exactly $a-1$ of $w_2^2$. For
the range  $a=Q+2,\dots ,2Q$ we do the analogous map to the state 
 $w_1^1w_1^2 \dots w_1^2w_2^2 \dots w_2^2 \theta_3^2 \theta_4^2$ but with  
$a-2-Q$ of $w_2^2$. They form the bosonic subspace and thus have
$(-1)^F=1$. Nonzero matrix elements can be determined by transforming the
scattering matrix into this basis. The diagonal ones can be expressed  
as 
\eq 
S^{sym}_{1-Q}(x^\pm,z^\pm)_{1i}^{1i}=\frac{Q+2-i}{Q+1}a^1_1(x^\pm,z^\pm)+
\frac{i-1}{2}a^2_2(x^\pm,z^\pm)\ \ ;\quad i=1,\dots, Q+1
\label{F1}
\eqx
\eq 
S^{sym}_{1-Q}(x^\pm,z^\pm)_{1i+Q+1}^{1i+Q+1}=\frac{2(Q-i)}{(Q-1)^2}a^4_4(x^\pm,z^\pm)+
\frac{i-1}{2}a^{10}_{10}(x^\pm,z^\pm)\ \ ;\quad i=1,\dots, Q-1
\eqx
while the non-diagonal ones are determined by 
\eq 
S^{sym}_{1-Q}(x^\pm,z^\pm)_{1i}^{1i+Q}=\frac{(i-1)(Q+1-i)}{(Q-1)}
a^4_2(x^\pm,z^\pm) \ \ ;\quad i=2,\dots, Q
\eqx
\eq 
S^{sym}_{1-Q}(x^\pm,z^\pm)_{1i+Q+1}^{1i+1}=\frac{1}{Q-1}a^2_4(x^\pm,z^\pm)
\ \ ;\quad i=1,\dots, Q-1
\eqx
For $a=2Q+1,\dots,3Q$ and  $a=3Q+1,\dots,4Q$ we choose the basis as 
 $w_1^1w_1^2 \dots w_1^2w_2^2 \dots w_2^2 \theta_3^2 $ and  
$w_1^1w_1^2 \dots w_1^2w_2^2 \dots w_2^2 \theta_4^2$ with $a-1-2Q$ and  
$a-1-3Q$ of $w_2^2$, respectively. They form the fermionic basis, thus,
corresponds to $(-1)^F=-1$, and have diagonal matrix elements only: 
\eq 
S^{sym}_{1-Q}(x^\pm,z^\pm)_{1i+2Q}^{1i+2Q}=\frac{Q+1-i}{Q^2}a^6_6(x^\pm,z^\pm)+
\frac{i-1}{2}a^7_7(x^\pm,z^\pm)\ \ ;\quad i=1,\dots, Q
\label{F2}
\eqx
The other fermionic space gives the same contribution
\eq 
S^{sym}_{1-Q}(x^\pm,z^\pm)_{1i+3Q}^{1i+3Q}=
S^{sym}_{1-Q}(x^\pm,z^\pm)_{1i+2Q}^{1i+2Q}\ \ ;\quad i=1,\dots, Q
\eqx
Putting all these things together gives the matrix part
$S^{matrix,su(2)}_{1-Q}(x^\pm, z^\pm)$. 

Let us use this result for the computation of the
relevant forward scattering amplitudes appearing in the L\"uscher type 
correction. In the case of the symmetric $Q-1$ scattering 
we have to make the $+\leftrightarrow -$  substitution:
\eq
FSA_{Q-1}^{sym}(z^\pm,x_i^\pm,x_{ii}^\pm)=FSA_{1-Q}^{sym}(x_i^\mp,x_{ii}^\mp,z^\mp)
\eqx
In the case of the antisymmetric representation and the $sl(2)$ sector we have
to replace the $su(2)$ scalar part with the $sl(2)$ scalar part, while for the
matrix part we have to take
\eq
S^{matrix,sl(2)}_{Q-1}(z^\pm, x^\pm)=S^{matrix,su(2)}_{Q-1}(z^\mp, x^\mp)
\eqx

\section{The case of symmetric representations}

In this appendix we would like to present the results of an analogous computation using the family of bound states which contains the \suii bound states. These states lie in the symmetric representation of each \sutt factor.

The scalar factor (\ref{e.suiiscalar}) for the forward scattering with the \suii bound state takes the form
\eq
S_{Q-1}^{scalar,su(2)}=\frac{3 \left(3 (q+i Q)^2+4\right) (q+i Q)^2+16}{3
   \left(3 (q-i Q)^2+4\right) (q-i Q)^2+16}
\eqx
In the above formula there should also be `string frame' phase factors (see \cite{ZF,AF1}). The ones involving the momenta of the two fundamental magnons which form the Konishi state cancel each other, while the ones involving the bound state combine with the $(z^-/z^+)^2$ factor to produce an effective length $L=4$ as expected from the spin chain perspective. Thus the effective exponential factor is
\eq
\left( \f{z^-}{z^+} \right)^4=\f{256\, g^8}{(q^2+Q^2)^4}
\eqx
Finally the matrix part can be evaluated to
\eq
S^{matrix,su(2)}_{Q-1}=\frac{576\, Q^2}{\left(3 q^2+6 i Q q+6 i q-3 Q^2-6
   Q-4\right)^2}
\eqx

\begin{figure}[t]
\begin{align}
&\Delta E^{su(2)}=-7776-9720 \gamma +1080 \pi ^2+48 \pi ^4
-6048 \zeta (3)-1440 \zeta (5)+ \nonumber\\
&+
\left(-7776+648
   i \sqrt{3}\right) \psi
   ^{(0)}\left(\frac{1}{2}-\frac{i}{2
   \sqrt{3}}\right)+
\left(-7776-648 i \sqrt{3}\right)
   \psi ^{(0)}\left(\frac{1}{2}+
\frac{i}{2
   \sqrt{3}}\right)+
\nonumber\\
&+
\left(2916-324 i \sqrt{3}\right) \psi
   ^{(0)}\left(1-\frac{i}{\sqrt{3}}\right)+\left(2916+324
   i \sqrt{3}\right) \psi
   ^{(0)}\left(1+\frac{i}{\sqrt{3}}\right)+\nonumber\\
&-\left(3240-540
   i \sqrt{3}\right) \psi
   ^{(1)}\left(\frac{1}{2}-\frac{i}{2
   \sqrt{3}}\right)-\left(3240+540 i \sqrt{3}\right) \psi
   ^{(1)}\left(\frac{1}{2}+\frac{i}{2
   \sqrt{3}}\right)+
\nonumber\\
&+
\frac{1}{2} \left(-1512+432 i
   \sqrt{3}\right) \psi
   ^{(2)}\left(\frac{1}{2}-\frac{i}{2
   \sqrt{3}}\right)+\frac{1}{2} \left(-1512-432 i
   \sqrt{3}\right) \psi
   ^{(2)}\left(\frac{1}{2}+\frac{i}{2
   \sqrt{3}}\right)+\nonumber\\
&-\frac{1}{6} \left(216-216 i
   \sqrt{3}\right) \psi
   ^{(3)}\left(\frac{1}{2}-\frac{i}{2
   \sqrt{3}}\right)-\frac{1}{6} \left(216+216 i
   \sqrt{3}\right) \psi
   ^{(3)}\left(\frac{1}{2}+\frac{i}{2
   \sqrt{3}}\right) \nonumber
\end{align}

\caption{The result for the symmetric representation bound states.}
\end{figure}
Again the resulting integral can be evaluated by residues with the result
\begin{align}
&\Delta E_Q^{su(2)}=-\frac{7776}{Q}+\frac{6480}{Q^2}-\frac{6048}{Q^3
   }+\frac{4320}{Q^4}-\frac{1440}{Q^5}-\frac{1944}{Q+1} -\frac{1944 (Q+1)}{3
   Q^2+1} \nonumber\\
&-\frac{15552 (Q+1)}{3 Q^2+6 Q+4}+\frac{324 (6 Q+1)}{3
   Q^2-3 Q+1}+\frac{324 (138 Q+11)}{3 Q^2+3
   Q+1}-\frac{972 (5 Q-4)}{\left(3 Q^2-3
   Q+1\right)^2} \nonumber\\
&+\frac{972 (23 Q+28)}{\left(3 Q^2+3
   Q+1\right)^2}+\frac{648 (3 Q-4)}{\left(3 Q^2-3
   Q+1\right)^3}-\frac{648 (81 Q+32)}{\left(3 Q^2+3
   Q+1\right)^3}  \nonumber\\
    & +\frac{324}{\left(3 Q^2-3    Q+1\right)^4}
+\frac{324 (36 Q+11)}{\left(3 Q^2+3Q+1\right)^4}
\end{align}
This expression can be summed over $Q$ using Maple. Although we see the appearance of $-1440 \zeta(5)$, the remaining part of the answer has a very complicated transcendentality structure involving higher $\psi$ functions evaluated for imaginary {\em irrational} arguments (see figure~1) and it seems that it cannot be recast in the form of a linear combination of $\zeta$ functions of odd arguments which seems to be required by gauge theory perturbative computations. Comparing this result with the simplicity of the answer which used the antisymmetric \slii bound states of the mirror theory we conclude that it is the physicality condition in the mirror theory which is relevant for the L{\"u}scher terms.

\end{document}